\documentclass[12pt,preprint]{aastex}
\usepackage{natbib}
\pdfoutput=1

\newcommand{\Ok}{\Omega_{\rm k} }

\newcommand{\mybf}{}

\newcommand{\mum}{\mu_{\rm m} }

\usepackage{graphicx}

\begin{document}

\author{A. Dorodnitsyn \altaffilmark{1,2},  
	    G.S. Bisnovatyi-Kogan \altaffilmark{3}, 
	    T. Kallman \altaffilmark{1}}

\altaffiltext{1}{Laboratory for High Energy Astrophysics, NASA Goddard Space Flight Center, Code 662, Greenbelt, MD, 20771, USA}
\altaffiltext{2}{Department of Astronomy/CRESST, University of Maryland, College Park, MD 20742, USA}
\altaffiltext{3}{Space Research Institute, 84/32, Profsoyuznaya st., Moscow, Russia}

\title{
AGN obscuration through dusty infrared dominated flows. I. Radiation-hydrodynamics solution for 
the wind.
}

\begin{abstract}
We construct a radiation-hydrodynamics model for the obscuring toroidal structure in active galactic nuclei.
In this model the obscuration is produced at parsec scale by a dense, dusty wind which is supported by infrared radiation pressure on dust grains.
To find the distribution of radiation pressure,
we numerically solve the 2D radiation transfer problem in a flux limited diffusion approximation. 

We iteratively couple the solution with calculations of
stationary 1D models for the wind, and obtain the z-component of the velocity. 

Our results demonstrate that for  AGN luminosities greater than $0.1 \,L_{\rm edd}$ external illumination 
can support a geometrically thick obscuration via outflows driven by infrared radiation pressure.
The terminal velocity of marginally Compton-thin models ($0.2<\tau_{\rm T}<0.6$), is comparable to or greater than the escape velocity.
In Compton thick models
the maximum value of the vertical component of the velocity is lower than the escape velocity, suggesting that a
significant part of our torus is in the form of failed
wind.  

The results demonstrate that obscuration via normal or failed infrared-driven winds is a viable option
for the AGN torus problem and AGN unification models. Such winds can also provide an important channel for
AGN feedback.
\end{abstract}

\section{Introduction}
The active galactic nucleus (AGN) unification scheme envisages the presence of a geometrically and optically thick, torus-like structure 
which wraps and hides a supermassive black hole (BH) and active parts of an accretion disk.
The paradigm relies on the property of such a structure to obscure the central regions of AGN in type II objects, making the torus responsible for the apparent dichotomy of active galaxies (e.g. \citet{AntonucciMiller1985}).

Direct evidence for the existence of the toroidal obscuration comes from
interferometric mid-infrared observations of the nearby Seyfert II galaxies such as NGC 1068, \citep{Jaffe2004}, and the Circinus
galaxy \citep{Tristram07}.  Studies such as these support the idea of a cold (T=100 -- 1000 K) torus situated 
approximately 1 pc away from a supermassive BH.  
These observations also reveal
the inner hot ($\sim 800$ K) funnel of the torus and the outer, colder ($\sim 300$ K) dusty component \citep{Raban09, Bock2000}.
Theoretical modeling \citep{KrolikBegelman86,Dorodnitsyn08b} also predicts that the torus funnel is significantly hotter than the rest of the torus body due to heating by X-rays generated in the inner parts of an accretion disk.

Indirect evidence for the geometrically thick obscuring structure located at parsec scales comes from observations of warm absorber gas.
Such observations of nearby Seyfert I galaxies by the grating
spectrographs on the X-ray telescopes {\it Chandra} and {\it XMM-Newton}
reveal rich X-ray line spectra in the 0.1--10 keV range, 
which contain
numerous lines from ions such as Fe, Si, S, O, Mg, and Ne, broadened and blue-shifted by 100 -- 1000 $\rm km\, s^{-1}$.
These have been detected from approximately half of low-redshift AGN \citep{Halpern84, Kaspi02,Steenbrugge05, Reynolds97, McKernan07}.
Numerical modeling shows that if the cold gas of the torus is exposed to extensive X-ray heating then an evaporative flow is formed.  {\mybf Simulations suggest that this gas is producing the warm absorber spectrum \citep{Dorodnitsyn08b, Dorodnitsyn09, Chelouche08}.}

The wind scatters radiation from the accretion
disk and broad-line region toward the observer, giving rise
to polarized radiation flux observed in the optical and UV \citep{AntonucciMiller1985}, and
predicted by theoretical modeling to exist in X-rays in the $0.1-10$ keV range \citep{Dorodnitsyn2010}.

One of the major problems which must be addressed by a theory of AGN obscuration is how the torus resists collapse into a geometrically thin disk. If the torus is supported by rotation and gas pressure then the temperature of the gas should be of the order of the virial temperature 
$T_{\rm vir,g}=2.6 \times 10^{5}\,M_{6}/r_{\rm pc}$K, where $M_{6}$ is the BH mass in $10^{6}\,M_{\rm \odot}$, and $r_{\rm pc}$ is the distance in parsecs.
Clearly, such temperatures cannot be reconciled with the existence of dust.

One solution to this problem is that the pressure of infrared photons on dust prevents the vertical collapse of the torus and supports its geometrical thickness. 
Comparing the energy density of the X-ray and UV-photons, $3.44\times 10^{-5}\, M_{6}/r_{\rm pc}^{2}\,{\rm erg\,cm^{-3}} $,
(assuming that the black hole radiates at half of its Eddington luminosity and a 30\% covering fraction of the Compton thick portion of the torus)
with the energy density of infrared radiation, we obtain that a gas-dust temperature of a few$\times 100$K is required if all these X-ray and UV photons are converted to the infrared. A more elaborate treatment (see Section \ref{sect2}) shows that if the temperature of the torus is a fraction of
$T_{\rm vir, r}   \simeq527\,(\frac{n/10^{7}}{r_{\rm pc}})^{1/4}-938\,(\frac{n/10^{8}}{r_{\rm pc}})^{1/4} \quad{\rm K}$, where $n$ is the number density, the torus thickness will be maintained by radiation pressure. Here $T_{\rm vir, r}$ is another definition of the virial temperature based on the radiation energy density of black-body radiation in a radiation-dominated plasma.

Alternative scenarios assume obscuration either from a warped disk or via a
magnetically-driven accretion disk wind. The first scenario \citep{Phinney89} implies that
the transition accretion disk is locally geometrically thin but strongly warped \citep{Sanders89}.
In a magnetically-driven wind scenario \citep{KoniglKartje94},  the torus is identified with the outer regions of a dense hydromagnetic outflow.

All these models, including the one which we propose in this paper, describe the torus as consisting of tenuous plasma. 
Regardless of the mechanism for the obscuration, the gas of the torus is self-gravitating, susceptible to various instabilities, and so possibly clumped or/and in the form of clouds (e.g \cite{Elitzur08}).

A global solution requires
modeling of the radiatively supported torus via multi-dimensional and multi-group radiation hydrodynamics simulations including self-gravitation. 
To accurately treat all the macro- and micro-physical processes known to be involved 
is not computationally feasible.
Thus approximate numerical and analytical solutions are useful. Such 
a solution for a static rotating torus was found by \cite{Krolik07}. Making a number of assumptions, he was able to obtain a semi-analytic model which showed that a rotating, static, and geometrically thick torus can be supported  by infrared radiation pressure on dust grains.

Dust opacity, is typically a $\sim\times 10$ times greater than the electron Thomson opacity, and thus the critical luminosity at which IR radiation becomes dynamically important is much smaller than the Eddington luminosity:
$ L_{\rm c} \simeq 10^{-2}-10^{-1}\,L_{\rm edd}$. If the temperature of the gas becomes larger than $T_{\rm vir, r}$ then the radiation
pressure prevails over gravity, and a model should include global plasma motions.

In this paper we construct a model in which radiation pressure on dust grains not only supports the geometrical thickness of the torus but induces mass loss through infrared pressure driven winds.  
In our model, a ''torus'' is represented by an extended, dense and cold wind rather than by a static gravitationally bound torus. 
It is interesting that the physical conditions in such a wind resemble those in red super-giant stars (except for the rotation) where radiation from a static ''core'' supports an extended, slowly outflowing envelope \citep{BK-Dor99,BK-Dor01}. In such stars, the
outflowing wind is driven by radiation pressure in the continuum, including significant contribution coming from radiation pressure on dust.

In what follows,
we numerically solve the equations of radiation hydrodynamics which describe the infrared-driven wind.
Our solutions strongly support the concept of a dynamical torus: Compton-thick obscuration in which the structure is
determined by infrared-driven flows of a dusty plasma.

The plan of this paper is the following: we begin with basic assumptions underpinning our model in Section \ref{sect1};  the onset of an outflow is analyzed in Section \ref{sect2};  
in Section \ref{sect2D} we derive the equations of radiation hydrodynamics describing the torus, 
and discuss appropriate boundary conditions;
our numerical method is outlined in Section \ref{sect5}; and results are presented in Section \ref{results}.
The paper concludes with the discussion of major results, validity of approximations adopted and the relevance of our model to a physical picture of real AGN.

\section{Dusty torus supported by infrared pressure}\label{sect1}

A spherically-symmetric distribution of fully ionized plasma around a central mass
can be gravitationally bound if the luminosity of the central object is $L<L_{\rm edd}$, where
$L_{\rm edd}$ is the Eddington critical luminosity

\begin{equation}
L_{\rm edd}=\frac{4 \pi c G M_{\rm BH}}{\kappa_{\rm T}}= 1.26 \times 10^{45}\,M_{7}\mbox{,}
\label{Lc_electrons}
\end{equation}
where $\kappa_{\rm T}=0.4\,{\rm cm^{2}\,g^{-1}}$ is the Thomson opacity due to electron scattering, and $M_{7}=M_{\rm BH}/(10^{7}M_{\odot})$. 

The inner parts of an accretion disk around a black hole, where most of the accreting gas potential energy is dissipated, generate copious X-ray and UV radiation.
Exposure of the outer region of an accretion disk to such radiation can have a profound effect on 
its structure and dynamics. {\mybf In the following the dust opacity is denoted as $\kappa$.}
In the UV the opacity of a single dust grain is significantly greater than $\kappa_{\rm T}$: 
$\kappa^{\rm UV}_{\rm gr}\simeq 6\times 10^{3} \kappa_{\rm T}$), adopting dust grain sizes $0.025-0.25 \,\mu {\rm m}$ \citep{Mathis77},
and dust grain density of $2-3\, \rm g\,cm^{-3}$.
Assuming perfect coupling between the dust  and gas, and a dust to gas mass ratio, $50-100$
the critical luminosity for the dust-plasma mixture becomes significantly less than the Eddington
luminosity:

\begin{equation}
L^{\rm UV}_{\rm c,dust}\simeq5\times 10^{-4}-0.01\, L_{\rm edd}\mbox{.}
\label{Lc_UVdust}
\end{equation}
For instance, if a cold slab of plasma is exposed  to unattenuated X-ray and UV radiation, a significant part of such radiation will be absorbed and reprocessed into infrared in a thin ''photospheric'' layer of thickness,
$\delta{\it l}/R_{1\rm pc} \simeq 1.3 \times 10^{-3}\,n_{7}^{-1}$ (Hereafter, where appropriate we denote $y_{x}$ to represent a quantity
$y$ scaled in terms of $10^{x}$ units of the same quantity, $y$).  

In the infrared, 
the Rosseland mean opacity, $\kappa$ of dust in the temperature range $10^{2} -10^{3}$ K is approximately $10-30$ times larger than that of the electron Thomson opacities \citep{Semenov03}. The dust opacity determines the critical luminosity, 

\begin{equation}
L_{\rm c}=\frac{4 \pi c G M}{\kappa} \simeq (0.03 - 0.1)\,L_{\rm edd}\mbox{,}
\label{Le_ddington_dust}
\end{equation}
 
If $\Gamma = L/L_{\rm edd}>1$, a spherically-symmetrical distribution of dust would be promptly blown away from approximately the dust condensation radius, $r_{d}\simeq 0.3-1.5$pc, for typical luminosities of $10^{45} -10^{46}$erg $\rm s^{-1}$ \citep{Barvainis87,Phinney89}. 

However, the presence of an equatorial accreting flow changes the picture. 
As a result of the reprocessing of external X-ray and UV radiation the incoming accretion flow (which otherwise would be geometrically thin) is pumped up with IR radiation and becomes geometrically very thick.
For example, it has been shown that
a thin disk (thin torus) eventually puffs up due to reprocessing of the hard X-rays in 10-100-keV range \citep{Chang_etal07}.

As it becomes sufficiently fat, the torus intercepts significant fluxes of soft X-rays and UV. 
Reprocessing of this radiation to the infrared
domain further pumps the torus interior with infrared photons, which become a major driving force in supporting the torus against collapsing back into a thin disk state. 

In a dusty plasma, the total pressure consists of that of an ideal gas, $P_{\rm g}$ and that of radiation, $\Pi$
\begin{equation}
\label{Pressure_total}
P=P_{\rm g}+ \Pi\mbox{,}
\end{equation}
where
\begin{equation}
P_{\rm g}=\frac{1}{\mu_{\rm m}}\rho{\cal R} T, \qquad \Pi=a\,T^{4}/3\mbox{,}
\label{Equations_of_State1}
\end{equation}
and ${\cal R}=8.31\cdot 10^7({\rm erg\, K^{-1} \, g^{-1}})$ is the universal gas constant, $a=7.56\cdot 10^{-15}({\rm erg\, K^{-4} \, cm^{-3}})$ is the radiation density constant, and $\mu_{\rm m}$ is the mean molecular weight. 
Given the great variety of physical conditions in dusty molecular gas we set  $\mu_{\rm m}=1$ throughout this paper. For simplicity, we do not consider models which involve clumping, such as those of 
\cite{KrolikBegelman88,BeckertDuschl04,HonigBeckert07}, and
assume continuous distributions of dust and gas.

The relative importance of radiation pressure is described by the parameter 
$\displaystyle \beta=P_{\rm g}/P\simeq \left(10^{3} \,\frac{ T^{3}_{3} } {n_{7} } +1\right)^{-1}$. 
At densities $n=10^{7}{\rm cm}^{-3}$,
$P_{g}\sim \Pi$ at $T\sim 80$K, and the ratio $P_{\rm g}/\Pi$ rapidly decreases at higher $T$, 
becoming 0.33 at $100$K, and 0.04 at $200$K. 
In the regime we are interested in, a ${\rm few}\times 100 \lesssim T\lesssim 1000$K, so that $P_{\rm g}\ll \Pi$.
Thus, in this paper we neglect gradients of the gas pressure
in the calculation of the equilibrium and dynamics of matter.

The radiation energy density in the region we are concerned with is mostly determined by infrared radiation.
On the other hand one can completely ignore the contribution from the mass-density of radiation, 
$\rho_{\rm r}=a T^{4}/c^{2}\simeq 8.4\times 10^{-24}\, ({\rm g\, cm^{-3}})$
as it  is much smaller than the mass-density of the gas, 
$\rho\simeq 8.35 \times 10^{-18}\, n_{7}\,({\rm g\, cm^{-3}})$.

In the simplified model considered in this paper (Section \ref{sect2D}) we relax the condition of vertical  balance of the torus and treat it as a wind
driven by the radiation pressure on dust. Self-gravity and clumpiness are ignored altogether. The equatorial inflow is implied but not calculated.
It is also implied that this equatorial accretion inflow replenishes the gas lost in the outflow, but we do not attempt to model such a connection.
We assume the flow is axially symmetric. 
One of the integrals conserved along the flux surface (i.e. such a surface which embraces a constant mass flux) is the specific angular momentum, $l$,
(e.g. \cite{Beskin09}), and we assume the foot-points of the streamlines are located at the equator.
Solving the momentum equation along $z$, we take into account only the $v_{z}dv_{z}/dz$ component of the 
$\bf v\cdot\nabla\,v$ term of the equation of motion. 
The z-component of the radiation force is calculated from a 2D distribution of the radiation energy density, $E(z,R)$ which is obtained from the diffusion equation. The latter is solved numerically in 2D adopting the flux-limited diffusion approximation. 

\section{The onset of the radiation driven wind}\label{sect2}
Before embarking on numerical calculations (Section \ref{sect2D}), it is instructive to consider a static model of a rotating torus. We approximate it by a spherically-symmetric distribution of plasma and radiation occupying a wedge of an opening angle $\theta_{0}$, and extending along spherical radial coordinate $r$. Local thermodynamical equilibrium is assumed throughout the torus. Let us assume, (in this section only) that the radiation flux $F$ which is given at the inner edge of such torus diffuses along $r$, as 
$L/(4\pi r^{2})\simeq -D_{T}\,dT/dr$ , where $L$ is the total luminosity, and $D_{T}=4a c T^{3}/(3\kappa\rho)$ is the diffusion coefficient. 
Limitations of such a model are obvious but 
for crude estimates we assume that there is no departure from spherical symmetry and $L$ is conserved. This model resembles that of
\cite{BKZeld68} who analyzed the onset of the stellar wind driven by high atmospheric opacity in the case of a
non-rotating star. Here we extend their analysis by adding rotation.

The onset of the wind can be approximately derived by considering 
the radial balance equation and the equation for $dT/dr$, at {\it fixed} $\theta$:

\begin{eqnarray}
\frac{1}{\rho}\frac{dP}{dr} &=&-\frac{G M}{r^{2}} + \left( \frac{l^{2}}{r^{3} \sin^{2}\theta }\right)\mbox{,}\label{PressureBalanceEquationStatic}\\
\frac{d \, T^{4}}{dr} &=& -\frac{3\kappa \rho}{a c} \frac{L}{4\pi r^{2}}\mbox{,}\label{dT4drStatic}
\end{eqnarray}
where $\theta$ is an angle measured from the vertical axis $z$; 
$R$ is the cylindrical radius, $l=\Omega R^{2}$ is the specific angular momentum, $\Omega(R)$ is the angular velocity which we assume to be constant on cylinders of constant $R$.
Dividing (\ref{PressureBalanceEquationStatic}) over (\ref{dT4drStatic}), and formally integrating at a fixed $\theta$ over $r$, and applying boundary conditions 
$P=0$, $T=0$ at the torus boundary, we obtain:

\begin{equation}
P=\frac{a}{3\Gamma_{\rm c}}\left(T^{4} -\int_{0}^{T}\,\frac{l^{2}}{l^{2}_{\rm k}\sin^{2}\theta} \, d T^{4} \right) \simeq
\frac{E}{3\Gamma_{\rm c}} \left( 1 - \frac{l^{2}/l^{2}_{\rm k}}{\sin^{2}\theta}  \right)\mbox{,}
\label{PressureStatic}
\end{equation}
where $l_{\rm k}= \sqrt{r G M}$ is the Keplerian specific angular momentum, and $\Gamma_{\rm c} = L/L_{\rm c}$, 
and $E=aT^{4}$ is the radiation energy density. 
To perform the integration in (\ref{PressureStatic}), we took into account that
$P$ is a single argument function of $r$, and assumed that 
$l^{2}/ (l^{2}_{\rm k} ) = \alpha^{2} = const$, i.e. a radial model is specified by $\theta$ and $\alpha$. 
Taking into account that ${\displaystyle \rho=\left( P-aT^{4}/3\right) \frac{\mum}{{\cal R} T}  }$, from (\ref{PressureStatic}), we obtain

\begin{equation}
\rho/T^{3} = \frac{a\mum}{3 {\cal R} \Gamma_{\rm c}} \left( 1- \Gamma_{\rm c} - \alpha^{2}/\sin^{2}\theta \right)\mbox{.}
\label{rho_over_T3}
\end{equation}
Substituting (\ref{rho_over_T3}) into (\ref{dT4drStatic}), and integrating, we obtain

\begin{equation}
T= T_{0} + \frac{\mu\,G M}{4\cal R}(1-\Gamma_{\rm c} -\alpha^{2}/\sin^{2}\theta) \left( \frac{1}{r} - \frac{1}{r_{0}}\right)
\label{T_r_Static}
\end{equation}
Where $T_{0}=T(r_{0})$ at some fiducial $r_{0}$. From equation (\ref{rho_over_T3}) we conclude that the specific entropy of the radiation-dominated gas,
$S=4/3\,a\, T^{3}/\rho$ is constant at constant $\theta$.
In the absence of rotation, equations (\ref{PressureStatic}), (\ref{rho_over_T3}), and (\ref{T_r_Static}) are reduced to the corresponding equations of \cite{BKZeld68}.  
In order for (\ref{rho_over_T3}) to be meaningful, the condition for a static torus follows:

\begin{equation}\label{T_r_StaticL}
\Gamma_{\rm c}\leqslant 1-\alpha^{2}/ \sin^{2}\theta \mbox{.}
\end{equation}
Alternatively, at fixed $\theta$, an outflow begins if the condition (\ref{T_r_StaticL}) breaks down.
From (\ref{T_r_Static}) follows a critical angle $\theta_{\rm c}=\arcsin(\alpha/\sqrt{1-\Gamma_{\rm c}})$, such that at $\theta < \theta_{\rm c}$ a static configuration with $\alpha=const$ is not possible.  At the equator, to be static such a torus should be sub Keplerian. 

Even if $\Gamma_{\rm c} < 1-\alpha^{2}/\sin^{2}\theta$ from (\ref{T_r_Static}) and from the condition that $T$ should {\it not} be finite at infinity we get another condition for the absence of an outflow

\begin{equation}
T_{0} < \frac{T_{\rm vir,g}}{4}(1-\Gamma_{\rm c} -\alpha^{2}/\sin^{2}\theta)\mbox{,}
\label{T_r_StaticT}
\end{equation}
where 

\begin{equation}\label{Tvir_gas}
T_{\rm vir,g} = \frac{GM}{r}    \frac{\mu_{\rm m}}{\cal R}\simeq 2.11\times 10^{5}\frac{M_{6}}{r_{\rm pc}}\,{\rm K}\mbox{,}
\end{equation}
is the gas virial temperature.
If (\ref{T_r_StaticT}) is violated a wind will occur due to a combination of thermal and radiation driving.

Using (\ref{rho_over_T3}) in (\ref{T_r_Static}) and adopting 
a similar line of arguments one can deduce another useful condition for an outflow:

\begin{equation}\label{outflow_cond2}
T_{0}>T_{\rm vir, r}(r_{0})\,\Gamma_{\rm c}^{1/4}\mbox{,}
\end{equation}
where 

\begin{equation}\label{Tvir_rad}
T_{\rm vir, r} = \left(\frac{GM\rho}{a r}\right)^{1/4} \simeq 312\,(\frac{n_{5} M_{7}}{r_{\rm pc}})^{1/4}-
987\,(\frac{n_{7}M_{7}}{r_{\rm pc}})^{1/4} \quad{\rm K}\mbox{,}
\end{equation}
is the second definition of the virial temperature which replaces (\ref{Tvir_gas}) in the case of $\Pi\gg P_{\rm g}$.
In the radiatively-dominated torus, if condition (\ref{outflow_cond2}) is fulfilled then an outflow driven 
by the pressure of the radiation flux, $F = -D_{T}\,dT/dT$ begins.

{\mybf The relation for $T_{\rm vir, r}$, (\ref{Tvir_rad}) does not contain opacity. Notice that in the diffusion approximation, the radiation force, $g_{\rm rad} = \kappa F/c= \kappa \lambda \,dE/dl \sim
(1/\rho) \, dE/dl $, where $\lambda$ is the photon mean free path, $\lambda=1/(\kappa\rho)$. 
In the optically thick case $dE/dl \sim E/L\sim \sigma T_{\rm eff}^{4}/(c L)$, where $L\sim r$ is the size of the system.
From balancing $g_{\rm rad}$ and gravity $GM/r^{2}$, the scaling (\ref{Tvir_rad}) is obtained.

In the free-streaming limit $F =cE =\sigma T_{\rm eff}^{4}$, where $\sigma = a c/4$ is the Stefan-Boltzmann constant, 
and here $T_{\rm eff}$ is the temperature of the photosphere or of a layer where external radiation is converted to IR. 
Adopting the same line of arguments as in the optically thick case the following relation can be obtained:

\begin{equation}\label{Tvir_rad_flx}
T_{\rm vir, flx} = \left( \frac{4GM}{a r^{2} \kappa}\right)^{1/4} =292\, (\frac{M_{7}}{r^{2}_{\rm pc} \kappa_{10}})^{1/4} \quad {\rm K}\mbox{.}
\end{equation}
Note that
the same result can be obtained assuming $dE/dl \sim E/\lambda$, which is also valid in the vicinity of the conversion layer (i.e. in thermalization layer).  
In the optically thin case the radiation pressure is determined by the anisotropic radiation flux and in the optically thick case by the gradient of $E$, which is determined by the size of the system.
The two effective temperatures are connected by the relation:

\begin{equation}\label{Tvir_flx}
T_{\rm vir, flx} = \frac{1}{\tau_{\rm d} } \, T_{\rm vir, r}\mbox{,}
\end{equation}
where $\tau_{\rm d}=\kappa \rho \, r$ is the optical depth parameter. 

The effective temperature of the conversion layer is found from 
$\alpha \Gamma \,F_{\rm edd} =\sigma T_{\rm eff}^{4}$ (here $\Gamma$ is related to the total BH luminosity and Thomson opacity):

\begin{equation}\label{Teff00}
T_{\rm eff} =\left(4\gamma\Gamma\frac{GM}{\kappa_{\rm T} {a r^{2}}}\right)^{1/4} \simeq 463\,(\frac{\Gamma_{0.5}M_{7}} { r^{2}_{\rm pc}} )^{1/4}
\quad {\rm K} \mbox{,}
\end{equation}
where $\gamma\simeq 0.5$ is the fraction of the incident flux reemitted in the IR inside the torus.
}  

\section{2D + 1D model}\label{sect2D}

\noindent
Now we describe the ingredients of our numerical model for the radiation and gas flow.
Consider equations describing stationary, slowly ($v \ll c$) outflowing wind.
The equation of motion and the continuity equation read:
\begin{eqnarray}
&{\bf v\cdot\nabla\, v} &= {\bf G_{\rm IR}} - \nabla\Phi  \mbox{,}\label{Euler}\\
&\nabla\cdot({\rho {\bf v}}) &= 0\mbox{,}\label{ContEq1}
\end{eqnarray}
where 

\begin{equation}
{\bf G}_{\rm IR} = \frac{\bf F}{c}\kappa\mbox{,}
\end{equation}
is the radiation force, and $\Phi=-GM/(z^{2} + R^{2})^{1/2}$ is the gravitational potential. For simplicity, we do not differentiate between the Rosseland and flux mean opacities \citep{MihalasBookRadHydro}, and set 
dust opacity, $\kappa$ constant. 
We adopt a diffusion approximation which
connects the infrared radiation flux, ${\bf F}$ with the infrared radiation energy density $E$:

\begin{equation}\label{eq1}
{\bf F}=-\frac{c}{3\kappa\rho}\,\nabla E = - D \, \nabla E \mbox{.}
\end{equation}
Notice, that  (\ref{eq1}) is in the form of the Fick's diffusion law.
The diffusion coefficient is 

\begin{equation}
D= c\, \lambda\mbox{,}\label{DifCoef}\mbox{.}
\end{equation} 

In this paper we do not consider external heating by hard X-rays, and thus in the bulk 
of the flow we have:
\begin{equation}\label{eq2}
\nabla \cdot {\bf F} = 0\mbox{.}
\end{equation}

At small optical depths (i.e when $\rho\to 0$), the standard diffusion approximation breaks down: the mean free path, $\lambda \to \infty$,  and
$D\to \infty$, and $F\to \infty$ instead of $F\to c E$ as it should be in a free-streaming limit. To take into account regions
of $\tau<1$ we adopt the flux-limited diffusion approximation \citep{AlmeWilson74,Minerbo78,LevermorePomraning81}.
In the flux-limited diffusion approximation $\lambda$ is replaced by $\lambda^{*} = \lambda\,\Lambda$,
where $\Lambda$ is the flux limiter. The flux limiter we adopt is that of \cite{LevermorePomraning81}:

\begin{equation}\label{LP_FluxLim}
\Lambda = \frac{2+R_{\rm LP}}{6+3R_{\rm LP}+R_{\rm LP}^{2}}\mbox{,}
\end{equation}
where $R_{\rm LP}=\lambda\,|\nabla E |/E$. If $\tau\to 0$, then $R_{\rm LP}\to \infty$, and 
$|F| \sim c\,E$. In the optically thick limit $R_{\rm LP}\to 0$ and $\Lambda\to 1/3$.

Adopting cylindrical ${z,R}$ coordinates, and assuming axial symmetry ($\partial/\partial \phi \equiv 0$) in the $\phi$ direction
we numerically solve equation (\ref{eq2}) in two dimensions.

\begin{figure}[htp]
\includegraphics[width=220pt]{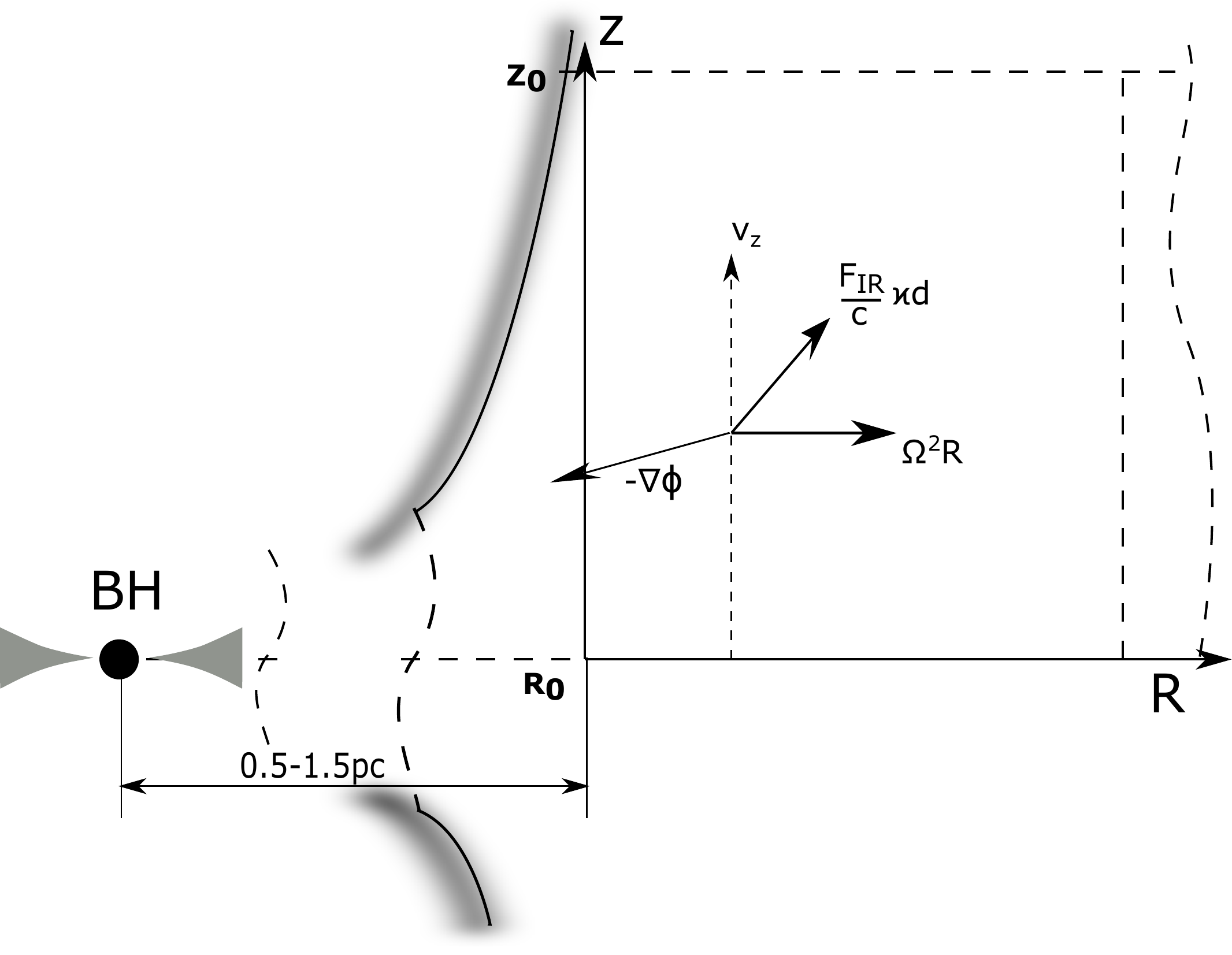}
\caption{Illustration of the flow geometry and a coordinate system implied by the
calculations (see the text for details). Not to scale.
}\label{sketch}
\end{figure}

In ${z,R}$ coordinates equation (\ref{Euler}) takes the form:

\begin{equation}
{\bf \hat e}_z(v \, {\partial_{z} v }   +z\,\Ok^2 )+ {\bf\hat e}_{R} (R\,\Ok^2- R\,\Omega^{2}) 
= - \frac{\kappa}{c} D \, \nabla E\mbox{,}\label{EqMotion2}
\end{equation}
where ${\bf \hat e}_z\,  {\bf\hat e}_{R}$ are coordinate unit vectors, and ${\partial_{x} y} \equiv \frac{\partial y}{\partial x}$. We allow for only the 
$v(z,R)_{z}$ component of the velocity (hereafter $v\equiv v_{z}$), i.e. an outflow is occurring along cylinders of constant $R$ (see the end of Section \ref{sect1} for discussion). The specific angular momentum, $l$ must be conserved along the flux surfaces: $l(R) = \Omega(R)\, R^{2}$, and thus the angular velocity, $\Omega$
is constant on cylinders of constant $R$. 
The geometry of the flow is shown in Figure \ref{sketch}.
The Keplerian angular velocity is found from

\begin{equation}
\Ok^{2}=\frac{GM_{\rm BH}}{(R^{2} + z^{2})^{3/2}}\mbox{.}
\end{equation}
Along the cylindrical flux surface, the continuity equation (\ref{ContEq1}) reduces to simply 

\begin{equation}
\rho\, v  = {\dot \mu}_{R}= const.\label{ContEq2}
\end{equation}
The amount of matter transported in $z$-direction, ${\dot \mu}_{R}$ is a function of the streamline, i.e. of $R$.
It is convenient to rewrite the above equations in dimensionless units: $x=R/R_{0}$, 
${\tilde z}=z/R_{0}$, ${\tilde E}=E/E_{0}$, ${\tilde \rho}=\rho/\rho_{0}$,
${\tilde v}=v/v_{0}$ (to simplify notation in the following we omit the tilde), where $R_{0}$, 
$E_{0}$, $\rho_{0}$, and $v_{0}$ are fiducial quantities.
 
From (\ref{EqMotion2}), the momentum equation
is cast in the following form:

\begin{equation}\label{EqMotion_z}
\rho v\,\frac{dv}{dz} = - \Lambda\,A_{1}A_{2} \,\partial_{z} E - A_{2} \, z\,\Ok^{2}\rho  \mbox{,}
\end{equation}
where we introduced dimensionless $\Ok^{2}=1/{(x^{2} + z^{2})^{3/2}}$, and the non-dimensional parameters, $A_{1}$, $A_{2}$:

\begin{equation}\label{param1}
A_{1}= \frac{E_{0}}{\Omega_{\rm k0}^{2} R_{0}^{2}\,\rho_{0}}\mbox{,} \quad A_{2}=\frac{\Omega_{\rm k0}^{2} R_{0}^{2}} {v_{0}^{2}}
\mbox{.}
\end{equation}
From (\ref{param1}) it follows that $A_{1}=T_{0}^{4}/T_{\rm vir,r}^{4}$, and $A_{2}=v_{\rm k0}^{2} /v_{0}^{2} $, where $v_{\rm k0}=(GM_{\rm BH}/R_{0})^{1/2}$, and $T_{\rm vir,r}$ is found from (\ref{Tvir_rad}).
 
In the current work we are not using the $R$-component of (\ref{EqMotion2}) and cannot deduce the distribution of $j$ in the moving wind self-consistently. That is because we are forcing matter to flow along cylinders of constant $R$ and the radiation flux is pushing from just one side. 
The numerical solution which we obtain in Section \ref{results} demonstrates that gravitation cannot balance centrifugal and radiation pressure forces and that departures from purely vertical motion should occur.

The non-dimensional continuity equation reads:

\begin{equation}
\mu = \rho v\mbox{,}\label{ContEq_nondim}
\end{equation}
where 
$\mu =  {\dot \mu}_{R}/(\rho_{0} v_{0})$, i.e. $\mu(R_{0})=1$.

In order to solve equations (\ref{eq1}), and (\ref{eq2}) we need to find the distribution of density, $\rho(z,x)$. This can be done solving equation of motion (\ref{EqMotion_z}) and then using (\ref{ContEq_nondim}).
It is convenient to convert equation (\ref{EqMotion_z}) directly to the equation for $\rho$, making use of (\ref{ContEq_nondim}),
and then numerically integrate this equation along the streamline. Thus, the equation for $\rho$ reads:

\begin{equation}\label{Eq_rho_z}
\frac{d\rho}{dz} = \frac{A_{2}}{\mu^{2}}\, \rho^{2}\left( \Lambda \, A_{1}\partial_{z}E +\rho\, z\Ok^{2} \right)\mbox{.}
\end{equation}
where $\partial_{z}E$ is known from the solution of the diffusion problem. Thus, equations (\ref{eq1}), (\ref{eq2}), (\ref{Eq_rho_z}) describe our problem.
We emphasize that, although our treatment of the gas dynamics is quasi-one-dimensional (i.e. a flow along cylinders), our treatment of the radiation is fully two-dimensional. 

\subsection{Boundary conditions and parameters governing the flow}
\label{BC}

We solve equations (\ref{eq1}), (\ref{eq2}), (\ref{Eq_rho_z}) numerically in cylindrical coordinates ${z,R}$. 
In these coordinates, the computational domain has a rectangular shape 
with one side spanning from $R_{0}$ to $R_{1}$, and the other from $z=0$ to $z_{1}$. 

At the left 
boundary, we specify the distribution of energy density

\begin{equation}\label{ELaw}
E(z,x_{0})  = E_{x0}  \,z^{-\epsilon} \mbox{,} \label{Elin}
\end{equation}

\noindent
{\mybf where $E_{x_{0}} = E(z=0, R_{0})/E_{0}$.
From the wind physics perspective,
the case of smaller $\epsilon$ mimics the situation when energy is deposited into the flow from the boundary over a longer region.

At the equatorial plane, at  $z=z_{0}$ the flux is calculated from:

\begin{equation}\label{bc-equator}
F_{z}(z_{0}, R) = -D \,dE/dz = \sigma T^{4}_{\rm eff}\mbox{,}
\end{equation}
where $T_{\rm eff}$ is calculated from
a ''photospheric'' boundary condition, i.e. when
$T_{\rm eff} = T(z_{0}, R)$ is obtained self-consistently when solving the 2D diffusion problem for $E$.

}

At the upper boundary we apply a free-streaming boundary condition: $|F|\simeq cE$. We tried several implementations of the boundary conditions at the right boundary to find that the
solution is not sensitive to their particular choice. 
However, it is reasonable to assume that the torus is close to being isothermal at larger $R$ and not too large $z$, and thus we pick ''zero flux'' boundary conditions at $R_{1}$.

In order to obtain the distribution of $\rho(z, R_i)$ on a particular flux surface, one needs to specify $\mu(R_{i})$. If we would have to match a stationary outflowing solution with a static solution in the accretion disk (i.e. vertical distribution of $\rho$) the situation would be equivalent to that described in \cite{BK-Dor99}: having at hand the vertical distribution of $\rho$ in the accretion disk one would smoothly match it with the corresponding wind solution.
This should be done at an arbitrary point $z_{0}$, provided $v_{z}(z_{0})\ll v_{s}(z_{0})$, where $v_{s}$ is the sound speed, and from that matching the unique value of $\mu$ would follow. In our case we specify $\rho(z=0,R)$, and we must also specify 
$\mu(R)$ (or $v(z=0, R)$). 
At the equatorial plane we specify power law distributions for $\rho$ and $\mu$:

\begin{equation}\label{DenLaw}
\rho(0, x) = x^{-d}\mbox{,} 
\end{equation}
We also choose that $\mu$ scales as density at the equator, $\mu(x) = x^{-d}$, to provide $v_{z}(z=0, x) = v_{0}$
is the same at all $x$ in the equatorial plane.

\section{Solution: outline of the method}\label{sect5}
Combining equations (\ref{eq1}) and (\ref{eq2}) we obtain the diffusion equation:

\begin{equation}\label{deq1}
\nabla \cdot {\bf F}=\frac{\partial}{\partial l_{1}}(D (\nabla E)_{z} ) + \frac{\partial}{\partial l_{2}}(x D (\nabla E)_{x} )\equiv
{\cal D}^{I}(E) + {\cal D}^{J}(E)=0\mbox{,}
\end{equation}
Equation (\ref{deq1}) is solved numerically adopting an alternative direction implicit scheme (ADI), i.e. \citep{Fletcher_book,
Fedorenko_book}.
Here we outline the method while the details are left to Appendix B.

The computational domain $\{z_{i}, x_{i}\}$, where $i=1,N_{i}$, and $j=1,N_{j}$, spans from $0$ to $z_{1}$, and from $x_{0}$ to $x_{1}$ respectively.
In our calculations, we adopt a $100\times100$ numerical grid which spans the $z= 0.1-2$ range in the $z$ direction, and the $x= 1-3$ range in the $x$ direction.
We make use of a staggered grid: quantities $E$, $\rho$, and $v$ are cell-centered, while $D$ is face centered, 
(c.f. \citep{TurnerStone01}).
In order to avoid approximation errors near the coordinate singularities when finite differencing in the curvilinear coordinates, we introduce volume elements $dl_{2} = x\,dx$ and $dl_{1}=dz$ ($dl_{1}=dz$ is introduced for consistency), e.g., \citep{StoneNorman92}.

In order to solve equation ($\ref{deq1}$), we introduce a pseudo time variable, $t$, and convert this equation into a time-dependent one:

\begin{equation}\label{deq2}
\partial_{t} E - {\cal D}^{1}(E) - {\cal D}^{2}(E)=0\mbox{,}
\end{equation}
where $\partial _{t} y$, ${\cal D}^{1}$, and ${\cal D}^{2}$ schematically represent finite difference operators over $t$ and along the alternative directions.

In the ADI scheme, a single time step from $t$ to $t+\delta \tau$ is made in the following manner: 
1) outer loop along $1^{\rm st}$ coordinate (for example), with half time-step $\delta\tau^{*}= \delta\tau/2$, with fully implicit scheme for the ${\cal D}^{1}(E)$. Schematically, we have:
$\partial_{t} ({E_{ij}^{*}}, E_{ij}) - {\cal D}^{1}(\vec E^{*}) - {\cal D}^{2}(\vec E)=0$, where 
$\vec E^{*}=(E^{*}_{i-1,j}, E^{*}_{i,j}, E^{*}_{i+1,j})$, and 
$\vec E=(E_{i,j-1}, E_{i,j}, E_{i,j+1})$, and
$\vec E^{*}=\vec E(t+ \delta\tau^{*})$,  i.e. 
applying a three-point stencil in a fully implicit numerical scheme for the update in 1 direction.
2)
Finally, iterating the outer loop in $2^{\rm nd}$ direction: $\partial_{t}({\hat E_{ij}}, E_{ij}^{*}) - {\cal D}^{1}(\vec E^{*}) - {\cal D}^{2}( \vec{\hat E})=0$, where 
$\hat E=(\hat E_{i,j-1}, \hat E_{i,j}, \hat E_{i,j+1})$, and
$\vec E=(E_{i-1,j}, E_{i,j}, E_{i+1,j})$, and obtaining ${\hat E}=E(t+ \delta\tau^{*})$.  

Diffusion coefficients are taken at the ''old'' time, which has a tremendous benefit compared to dealing with linearized equations as would otherwise  
be necessary (in a fully implicit method). 
The fully implicit approach to the solution of a flux-limited diffusion problem was taken, for example by \cite{Hayes06}.

As a consequence of a three-point finite differencing stencil implied by the diffusion operator in (\ref{deq1}), the corresponding
matrix equation for the updated $E^{*}$ and ${\hat E}$ involves a tri-diagonal matrix.
We adopt a sweep method in order to solve the resultant tri-diagonal matrix equation via a tridiagonal matrix algorithm \citep{Fedorenko_book}.

A finite difference representation of the boundary conditions (BC) is derived in a way that preserves 
$2^{d}$ order accuracy of the numerical scheme.
In the ADI method, one can apply a combination of flux and temperature BC {\citep{Fletcher_book}.
However the flux at the boundary should be parallel to one of the coordinate lines \citep{Fedorenko_book}. 
In our model the inner boundary is parallel to $z$ and the flux should be normal to that boundary. 
Given our ignorance of the structure of the conversion layer at the inner boundary, we believe that is is slightly more physical to specify temperature BC instead of the flux one.
Thus, for simplicity we specify the distribution of the effective temperature at the innermost cylinder, which marks the inner boundary of the computational domain.

After the distribution of the radiation energy density, $E(z,R)$ is obtained, the next approximation for $\rho(z,R)$ is found from 
(\ref{Eq_rho_z}). We solve this equation along cylinders, in a $z$ direction, adopting a $4^{\rm th}$ order Runge-Kutta method \citep{NRBook92}. 
The updated distribution of $\rho$ is used to compute diffusion coefficients from (\ref{DifCoef}) and again to solve (\ref{deq1}), etc. The cycle is repeated until a stationary wind solution is found.

\section{Results of the numerical model}\label{results}
{\mybf 
It has been shown that the effective temperature of the conversion layer scales approximately as
$463\,({\Gamma_{0.5}M_{7}} /{ r^{2}_{\rm pc}} )^{1/4}$K, which is greater 
than $T_{\rm vir,r}\simeq 312$K and $T_{\rm vir,flx}\simeq 292$K for a $M_{7}=1$ BH and $n_{0} =10^{5}$. 
It is reasonable to expect that no equilibrium is possible between radiation pressure and vertical component 
of gravity and that a dynamic, outflowing atmosphere is a better description of what is going on. 

Our boundary conditions do not provide optimum acceleration as the incident radiation is normal 
to the flow at the boundary. 
It is the readjustment of the radiation flux inside the torus that produces a vertical gradient of $E$.
Since in our simplified method we can calculate only the $v_{z}$ component of the velocity it is quite possible that taking into 
account the full 2D picture can increase terminal velocity ($\nabla E$ has its largest component approximately parallel to spherical $r$).
The parameter $n_{0}$ scales the wind loading density. This density can be significantly smaller than the density at the equatorial part of the accretion disc.

It is instructive to compare  
$v_{0}\simeq 210\, (M_{7}/R_{\rm pc})^{1/2}\, {\rm km\,s^{-1}}$, with
the sound velocity in the radiatively dominated plasma.
Notice that in a radiation-dominated plasma $v_{s}\simeq \sqrt{E/3\rho}$, i.e. its value explicitly depends on both density and temperature. 
For relevant parameters we obtain:
$v_{\rm s}\simeq v_{\rm s,rad}\simeq 177 (T_{3}^{4}/n_{7})^{1/2}\,{\rm km\,s^{-1}}$. 
{\mybf
It is important that the wind launching speed is subsonic, and we 
choose $v_{0}=0.1 v_{s}$ for all models. Note that since $v_{s}$ depends on $\rho(z=z_{0})$, $v_{0}$ depends on it as well. 

We parametrize our models by the Thomson optical depth of the torus, $\tau_{\rm T}=\int_{0}^{\infty}\, \kappa_{\rm T}\rho \,dR$ calculated at the inclination $90^{\circ}$
from the z-axis adopting the equatorial distribution of density (\ref{DenLaw}) with $d=0.5$ throughout all of the models.
At the left boundary we choose $E_{x0}=1$, and $\epsilon=0.1$ in (\ref{Elin}) and use $\Gamma$ as a parameter instead of $T_{\rm eff}$ which is calculated from (\ref{Teff00}).

} The mass of the black hole is $M_{\rm BH} =1\times 10^{7}\ M_{\odot}$, and $R_{0}=1$pc, and $\kappa=10$ are fixed for all models.

Our results and various parameters of the models are summarized in Table~1.
In the following we describe several characteristic models from the above set.

We are not able to calculate models for $\Gamma \lesssim 0.1$ due to the intrinsic incapability of our method to treat low-velocity, 
decelerated flows. Such models require a full time-dependent multi-dimensional, radiation-hydrodynamics treatment.

Models with the characteristic BH luminosity as low as $0.1L_{\rm edd}$ produce a noticeable wind provided the optical depth is not too high.  
At larger optical depths the characteristic temperature at the equator is too low.
As a result, we do not obtain an outflow solution for $\Gamma = 0.1$ and $\tau_{\rm T}\gtrsim 0.6$. 
Increasing the optical depth from $\tau_{\rm T}\simeq 0.2$ to  $\tau_{\rm T} \simeq 0.5$ doubles the mass-loss rate to approximately $2M_{\odot}\,{\rm yr^{-1}}$ but also reduces the  maximum velocity, $v_{\rm max}$ by a factor of two. Most of the gas 
does not reach $U_{\rm esc}$ forming a failed wind. With increasing $\tau_{\rm T}$ the kinetic luminosity drops by an order of magnitude to $L_{\rm kin}\simeq 1.2 \cdot 10^{38}{\rm (erg\,s^{-1}) }$, which is an order of magnitude smaller than that 
obtained from
simple estimates of the kinetic luminosity: $L_{\rm kin} \simeq {\dot M} v_{\rm max}^{2}/2 \simeq 9.7\cdot 10^{39}{\rm (erg\,s^{-1}) }$. This is because only a fraction of the domain is occupied by the fast wind.

\begin{tabular}{c c c c c c c c c c}
Model & $\Gamma$ & $R_{0}$ & $\tau_{\rm T}$ & $n_{0}$ &$v_{\rm max}$  & 
$L_{\rm kin}  $ & $L_{\rm bol} $ & ${\dot M} $\\
\hline
\hline
$1$ & 0.1 & 1 & 0.17& $1\cdot 10^{5}$ & 215 & $4.22\cdot 10^{39}$ & $1.24\cdot 10^{44}$ & 1.23\\
$2$ & 0.1 & 1& 0.34& $2\cdot 10^{5}$ & 153 & $2.24\cdot 10^{38}$& $1.24\cdot 10^{44}$ &1.74\\
$3$ & 0.1 & 1 & 0.51& $3\cdot 10^{5}$ &123 & $3.59\cdot 10^{38}$ & $1.24\cdot 10^{44}$ & 2.14\\
$4$ & 0.3 & 1& 0.17 & $1\cdot 10^{5}$ & 311 & $1.63\cdot 10^{40}$ & $3.74\cdot 10^{44}$ & 1.59\\ 
$5$ & 0.3 & 1& 0.34 & $2\cdot 10^{5}$ & 217 & $3.98\cdot 10^{39}$ & $3.74\cdot 10^{44}$ &  2.25\\ 
$6$ & 0.3 & 1& 0.51 & $3\cdot 10^{5}$ & 251 & $1.33\cdot 10^{39}$ & $3.74\cdot 10^{44}$ &  2.76\\ 
$7$ & 0.3 & 1& 0.85 & $5\cdot 10^{5}$ & 129 & $5.34\cdot 10^{38}$ & $3.74\cdot 10^{44}$ &  3.56\\ 
$8$ & 0.5 & 1& 0.17 & $1\cdot 10^{5}$ & 445 & $4.65\cdot 10^{40}$ & $6.24\cdot 10^{44}$ &  2.05\\ 
$9$ & 0.5 & 1& 0.51 & $3\cdot 10^{5}$ & 357 & $5.9\cdot 10^{40}$ & $6.24\cdot 10^{44}$ &  4.1\\ 
$10$ & 0.5 & 1& 1.48 & $5\cdot 10^{5}$ & 203 & $1.2\cdot 10^{40}$ & $6.24\cdot 10^{44}$ &  5.29\\ 
$11$ & 0.5 & 1& 2.37 & $8\cdot 10^{5}$ & 129 & $2.76\cdot 10^{39}$ & $6.24\cdot 10^{44}$ &  6.69\\ 
$12$ & 0.8 & 1& 1.48 & $5\cdot 10^{5}$ & 318 & $4.2\cdot 10^{40}$ & $9.99\cdot 10^{44}$ &  6.7\\ 
$13$ & 0.8 & 1& 2.37 & $8\cdot 10^{5}$ & 191 & $1.04\cdot 10^{40}$ & $9.99\cdot 10^{44}$ &  8.47\\ 
$14$ & 0.8 & 1& 2.97 & $1\cdot 10^{6}$ & 162 & $5.56\cdot 10^{39}$ & $9.99\cdot 10^{44}$ &  9.47\\
$15$ & 0.8 & 1& 5.94 & $2\cdot 10^{6}$ & 104 & $1.19\cdot 10^{39}$ & $9.99\cdot 10^{44}$ &  13.4\\ 
\hline
\end{tabular}

\tablename{ 1. Models characterized by different initial parameters: $\Gamma$, $R_{0}(\rm pc)$, $\tau_{\rm T}$, characteristic density $n_{0}{(\rm cm^{-3})}$, and resulting 
kinetic and bolometric luminosities, $L_{\rm kin}{\rm (erg\,s^{-1}) }$, $L_{\rm bol}{\rm (erg\,s^{-1}) }$, and mass-loss rates 
${\dot M} (M_{\rm \odot}\,{\rm yr^{-1}} )$.
}

The density and radiation energy density for Model 1 are shown in Figure \ref{fig1}, and Figure \ref{fig2} shows the 
surface plot of the velocity, $v/U_{\rm esc}$, where  $U_{\rm esc}$ is the local escape velocity. Recall the distributions of $E$ and $\rho$ at the appropriate boundaries (\ref{Elin}). The most appropriate conditions for the acceleration of the wind happen in the middle of the domain, where the radiation field has strong gradients, but density is lower than that at the left boundary. The maximum value of the effective temperature  $T_{0}=T(R_{0})=407$K which rapidly declines at larger spherical radii, $r$. Maximum velocity attained by the wind is $4.7 \mbox{\it Ma} $, where {\it Ma} is the Mach number. 
Further increasing $\tau_{\rm T}$
results in a drop of the velocity: Most of the wind has velocity smaller than escape velocity, however the ''mass-loss rate'' 
of such a failed wind is noticeably larger (c.f. Table 1).

From Figure \ref{fig1} (right panel) one can see a significant drop of radiation energy density, $E$ 
within the distance, $\delta x\simeq 1.5$ from the left boundary and
from Figure \ref{fig2} we identify the region $x \simeq 1.5 - 2.5$ and where the most of the fast wind is blowing. 

As was discussed in Section \ref{BC}, in the approach taken in this paper, the mass-loss rate from the two sides of the disc, ${\dot M}$ is a mere consequence of the adopted boundary conditions.
From the continuity equation (\ref{ContEq_nondim}) $\rho \sim \mu/v_{z}$, and thus an increase of the velocity in region I is compensated by the reduced density in accord with
what is observed in Figure \ref{fig1} (left).  The enhanced density region in Figure \ref{fig1} (left panel) corresponds to a 
low velocity, high density and quasi-isothermal region of the torus.
In the following we denote the high velocity part as region I, the higher density, narrow transition region as region II and the high-density region located at larger radii as region III.

\begin{figure}[htp]
\includegraphics[width=450pt]{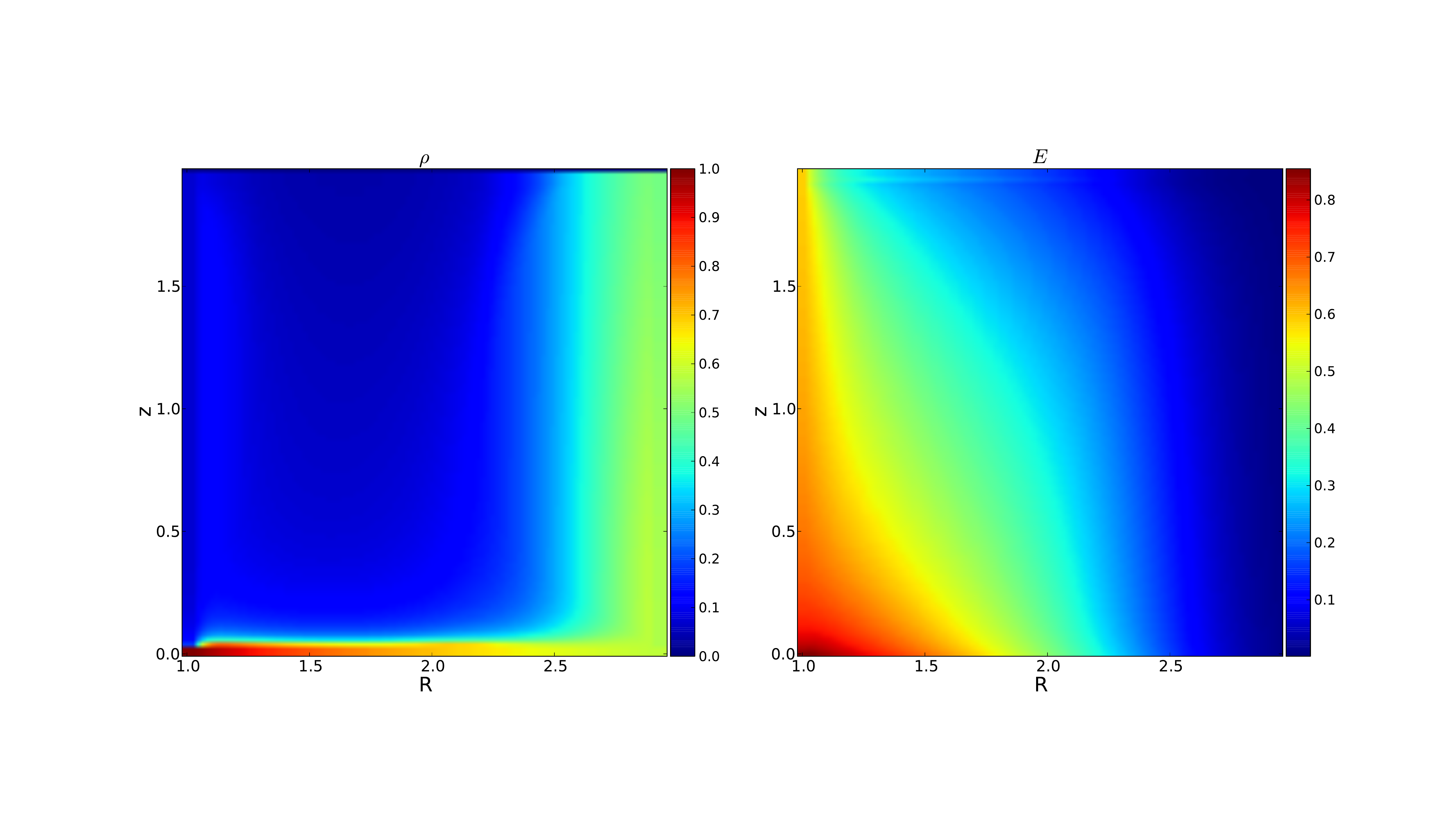}
\caption{Model 6: Color-intensity plots of the dimensionless density, $\rho$ (left) and dimensionless infrared radiation energy density, $E$ (right). Axes: distance in parsecs.
}\label{fig1}
\end{figure}

\begin{figure}[htp]
\includegraphics[width=220pt]{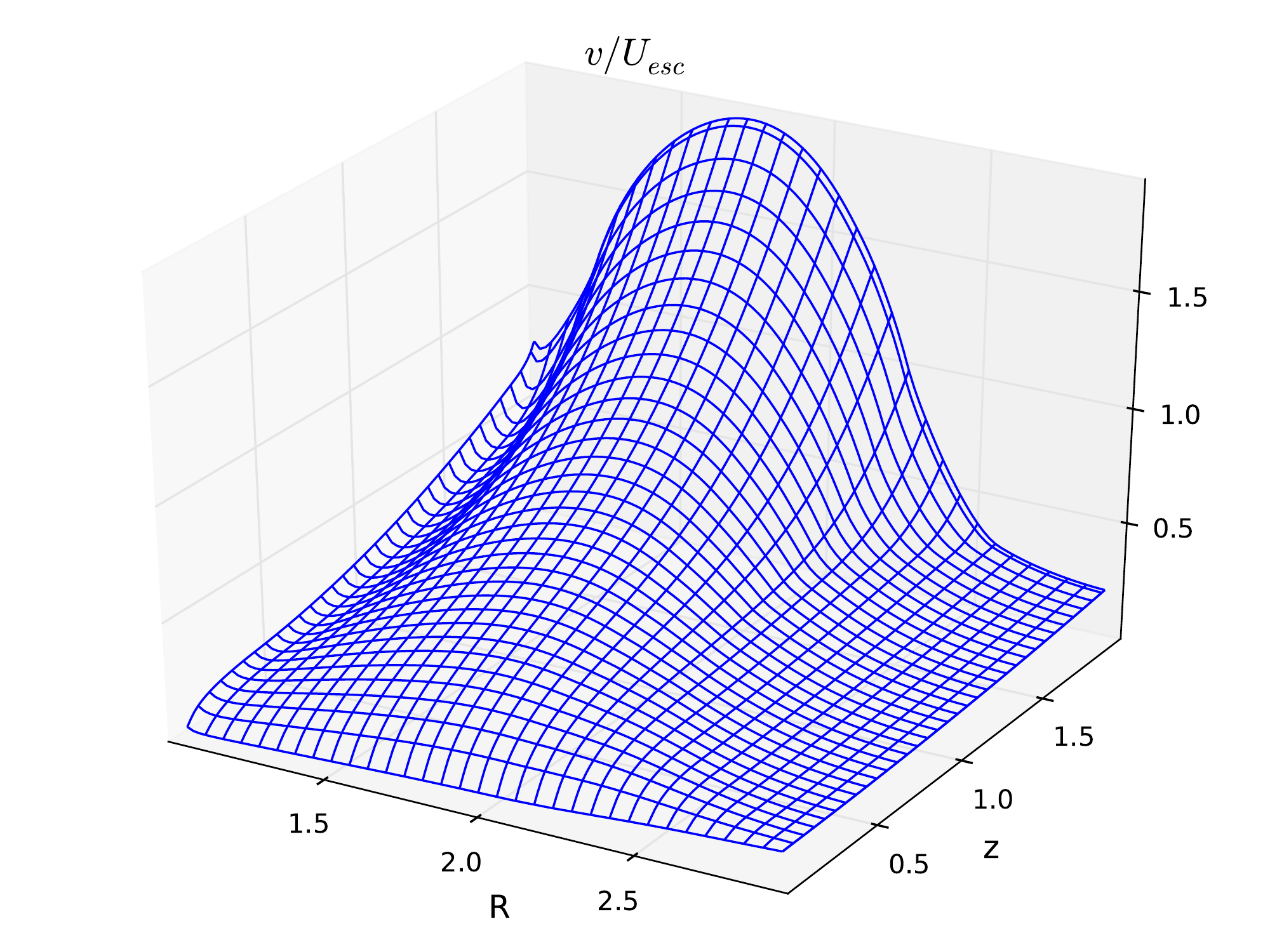}
\caption{Model 6: Velocity surface plot.  Axes: vertical: velocity $v_{z}/U_{\rm esc}$, where $U_{\rm esc}$
is the local escape velocity; horizontal: $R$: distance from the BH in parsecs;
$z$: distance from equatorial plane in parsecs;
}\label{fig2}
\end{figure}

{\mybf
Models 8-11 have $\Gamma=0.5$. These are optically thin, marginally optically thick and Compton thick models.
Increasing $\tau_{\rm T}$ from $\sim 0.2$ to $\sim 1.5$ results in increasing 
${\dot M}$ from  $\sim 2\,M_{\rm \odot}\,{\rm yr}^{-1}$ to $\sim 5.3\,M_{\rm \odot}\,{\rm yr}^{-1}$. 
The maximum velocity drops from $v_{\rm max}\simeq 445\rm\, km\, s^{-1}$ for Model 8, to 
$\sim 203\rm\, km\, s^{-1}$ for Model 10.
The color intensity plot of $\rho(z,x)$ and $E(z,x)$ are shown in Figure \ref{fig3}.
One can see the fast wind occupies approximately 40\% in the radial extent, and that in the 
wind region the density 
is markedly lower then in the outer parts. At larger $R$ there is no significant outflow. 
Notice the locus of a sharp rise of the density which marks a barrier between 
the region of fast flow and the almost quasi-static torus. The wind is supersonic, for example, the maximum
Mach number for Model 9 is 6.
}

\begin{figure}[htp]
\includegraphics[width=420pt]{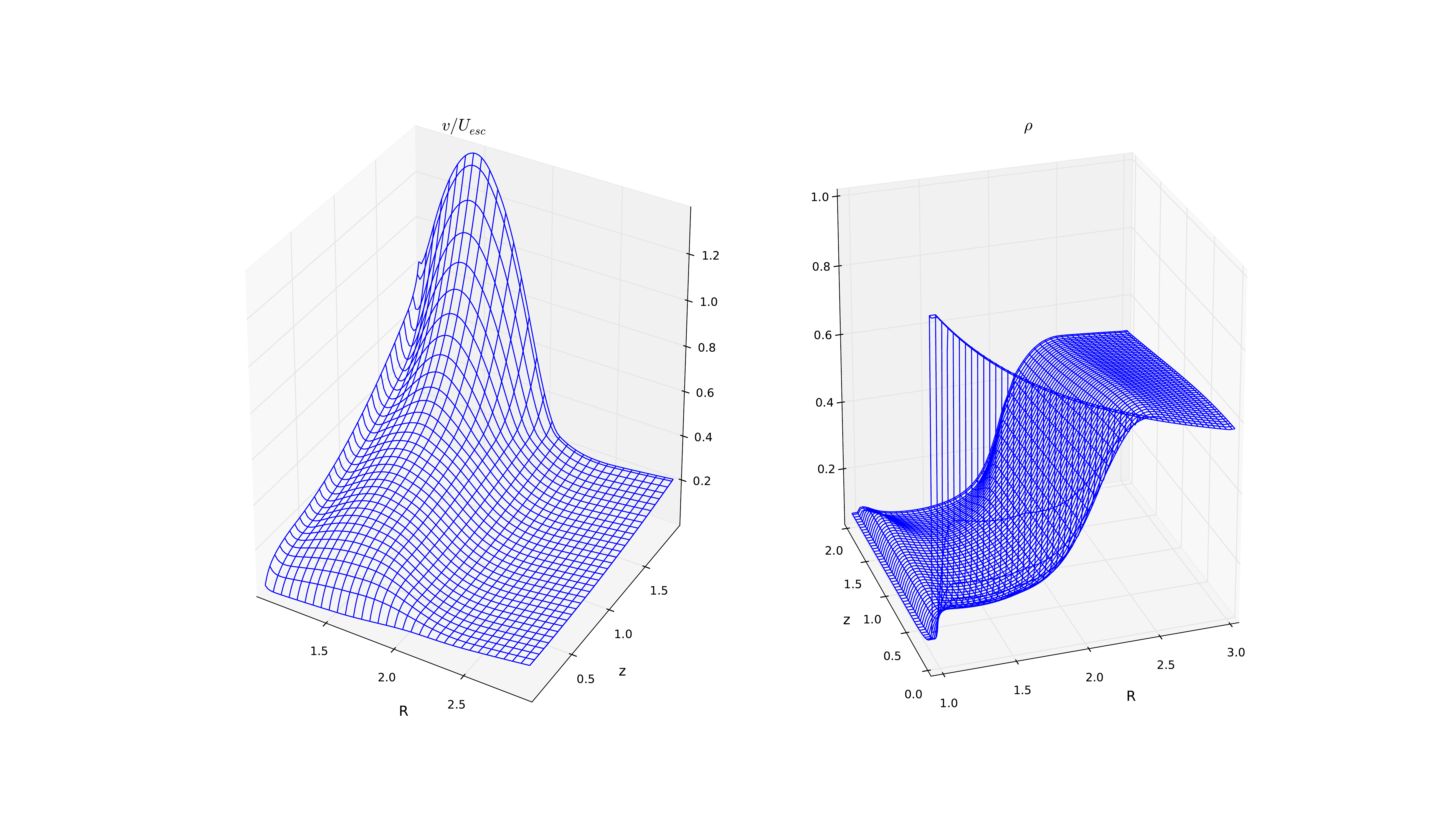}
\caption{Model 10: Left: velocity surface plot, where $v_{z}/U_{\rm esc}$, and $U_{\rm esc}$
is the local escape velocity. Right: density surface plot.  
Horizontal: $R$: distance from the BH in parsecs;
$z$: distance from equatorial plane in parsecs;
}\label{fig3}
\end{figure}

\begin{figure}[htp]
\includegraphics[width=450pt]{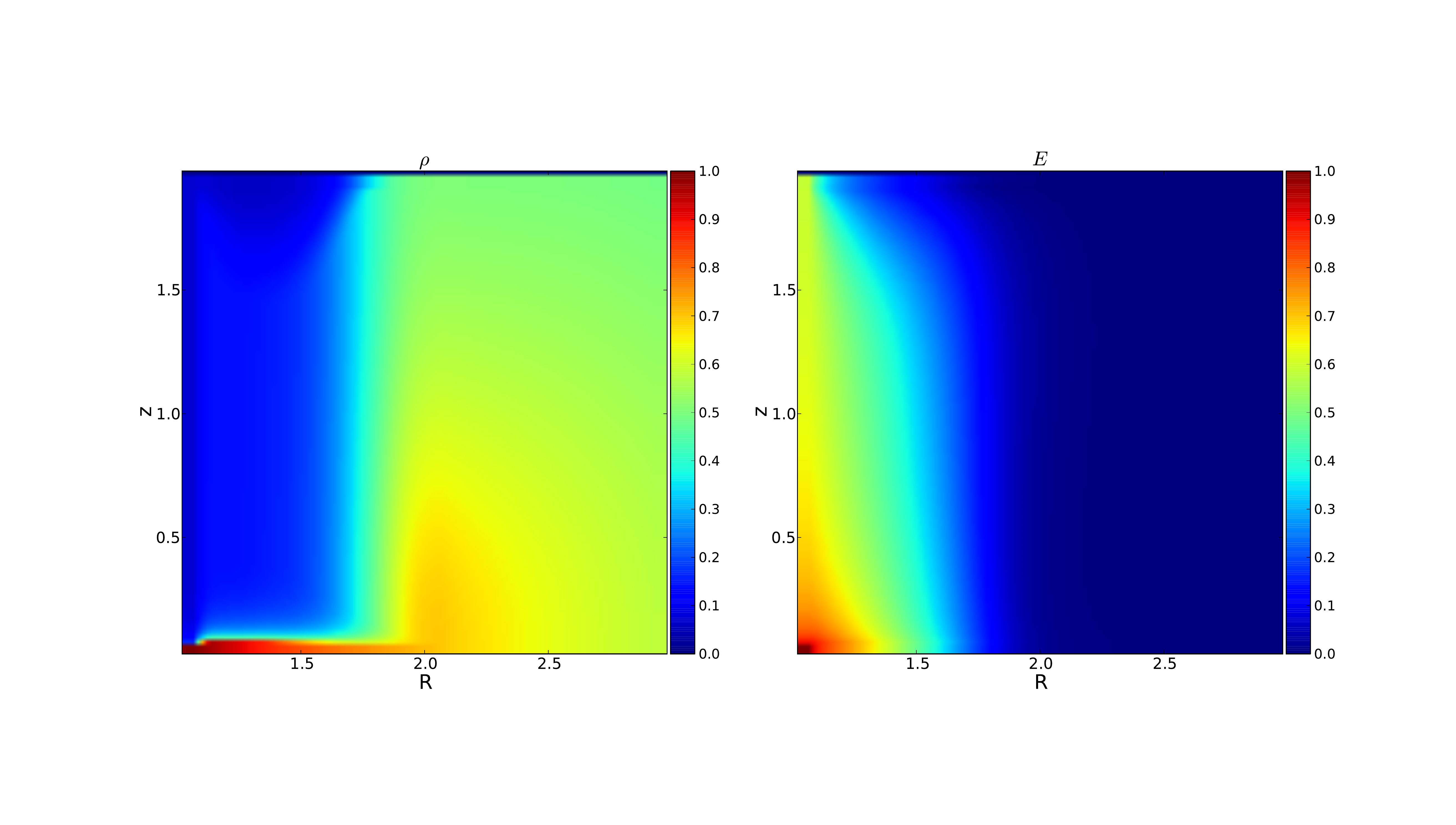}
\caption{Model 11. Color-intensity plots of the dimensionless density, $\rho$ (left) and dimensionless infrared radiation energy density, $E$ (right). Axes: distance in parsecs.
}\label{fig4}
\end{figure}

Further increase of $\tau_{\rm T}$ to 2.37 engages more matter into the low velocity wind.
The color intensity plots of $\rho(z,x)$ and $E(z,x)$ are shown in Figure \ref{fig4}. 
The wind in Model 11 does not reach the local escape velocity: the maximum velocity is 
$0.6\,U_{\rm esc}(z,R)$. Large amounts of gas, $6.69 M_{\rm \odot}\,{\rm yr}^{-1}$ 
participate in a low velocity $\sim 100 \rm\, km\, s^{-1}$ flow,
leaving the computational 
domain in the form of a failed wind. The high-density region III is clearly seen in Figure \ref{fig4} (left panel).

Models 12 -15 in Table 1 represent a BH shining close to $L_{\rm edd}$: They have $\Gamma = 0.8$;
These models are Compton-thick: $\tau_{\rm T} \simeq 1.5 - 6$. 
Results for Model 13 are shown in Figure~\ref{fig5}, and for Model~15 in Figure~\ref{fig6}.
From Figure~\ref{fig4} and Figure~\ref{fig6} one can see that within a high density region there is a higher density ''core'', which is most pronounced in Model 15.

\begin{figure}[htp]
\includegraphics[width=450pt]{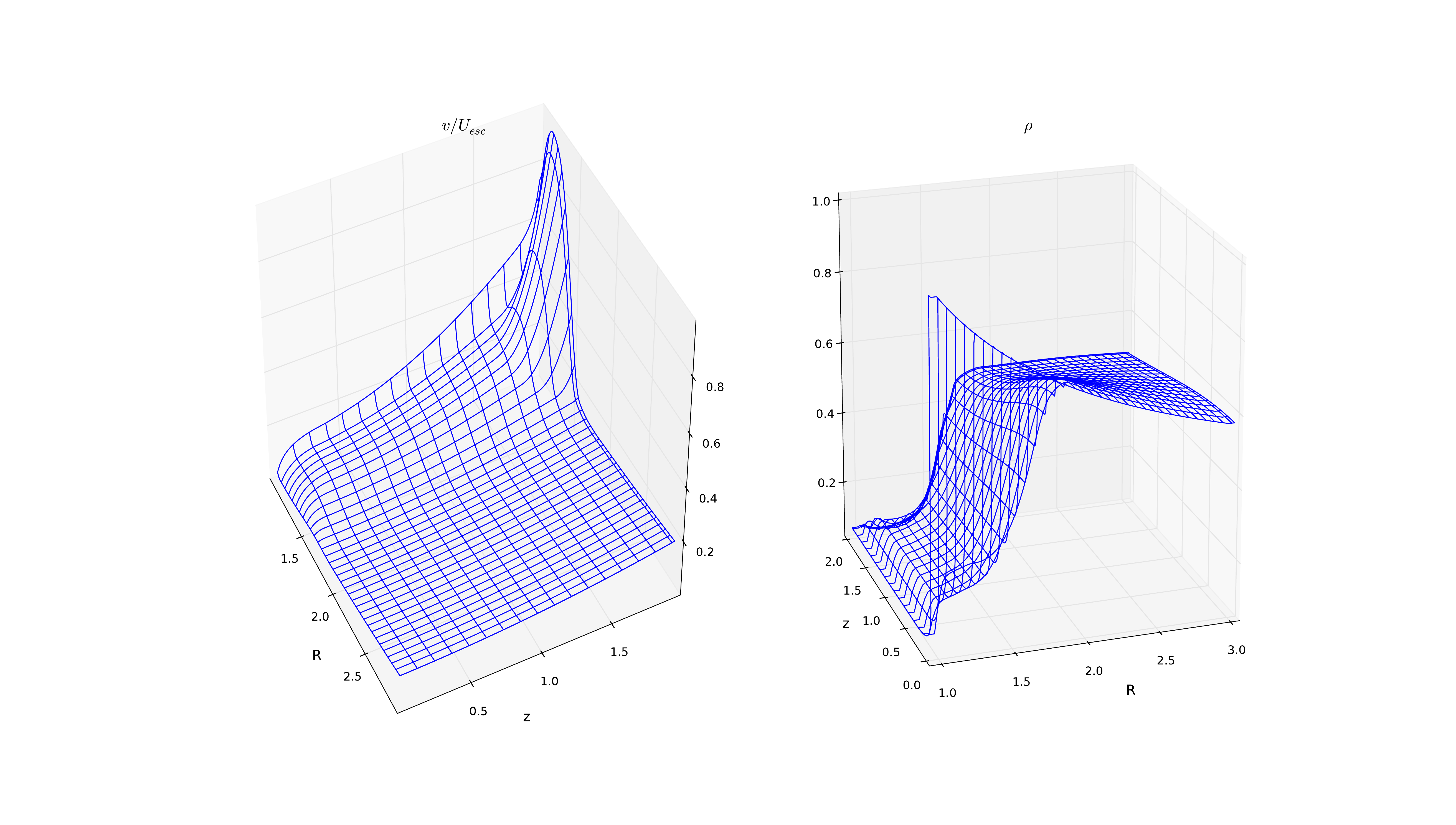}
\caption{Model $13$: 
Left: velocity surface plot. Right: density surface plot.
Axes: vertical: left: velocity $v_{z}/U_{\rm esc}$, where $U_{\rm esc}$
is the local escape velocity; right: density; horizontal: $z$: distance from equatorial plane in parsecs; $R$: distance from the BH in parsecs;
}\label{fig5}
\end{figure}

\begin{figure}[htp]
\includegraphics[width=230pt]{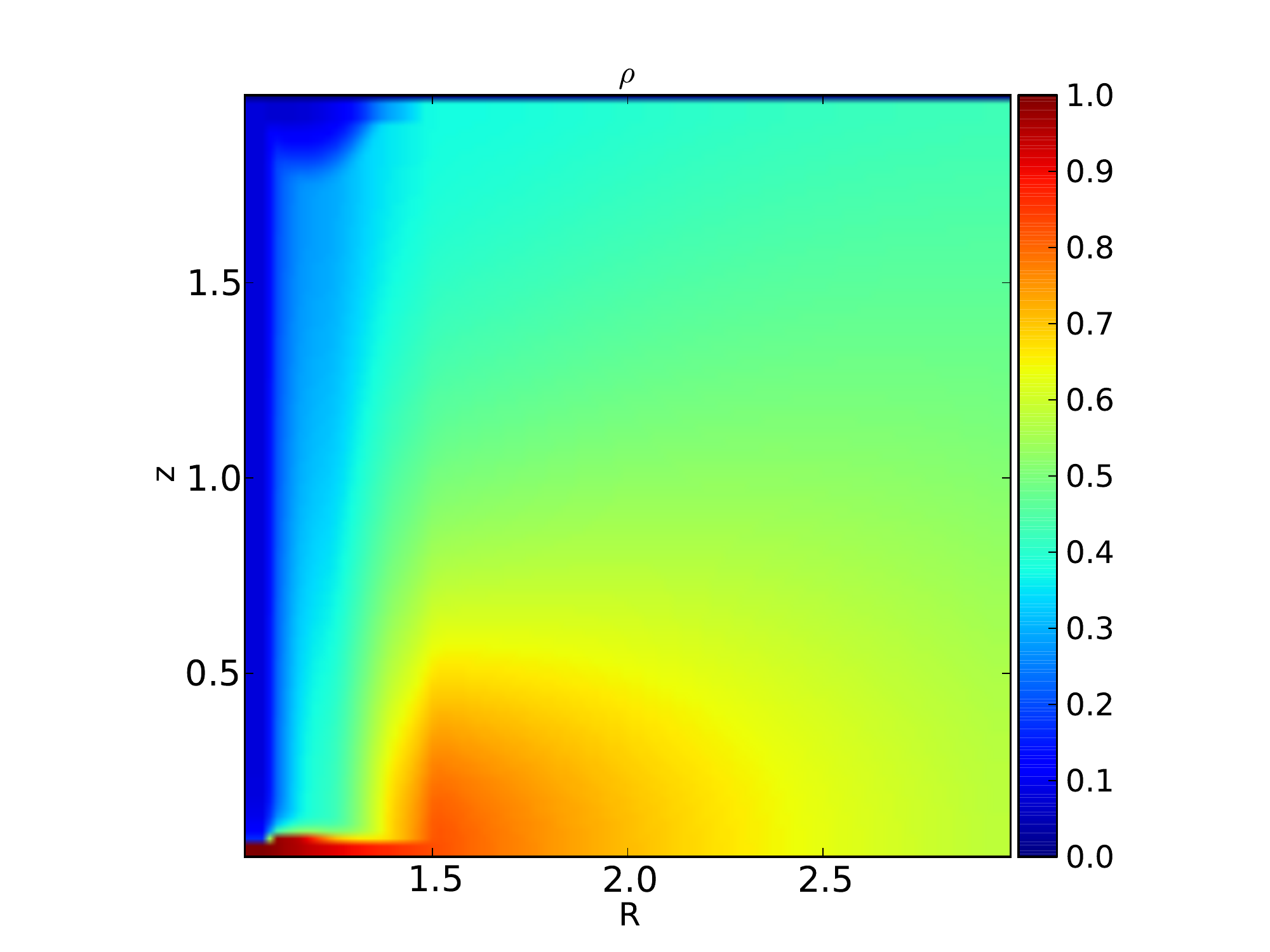}
\caption{Model 15. Color-intensity plots of the dimensionless density. Axes: distance in parsecs.
}\label{fig6}
\end{figure}

The high velocity wind in Model 15 occupies a narrow wedge-like region close to the left boundary, but even there,
the maximum velocity, $v_{\rm max}\simeq 0.5\, U_{\rm esc}(z,R)$. Low velocity of the wind translates into 
a ratio $L_{\rm kin}/L_{\rm bol}$ dropping to $\sim 8\cdot 10^{-7}$. Most of the flow is mildly supersonic,
{\it Ma}$ \,\simeq 1.5-2$ with velocity below the escape velocity. 
Here ${\dot M}$ is better interpreted not as a mass-loss rate but as a parameter describing how much gas is involved in large-scale motions; its value approaches $13.5\,M_{\rm \odot}\,{\rm yr}^{-1}$.

Most of our models demonstrate a clear separation of the torus into 
lower and higher density parts. From arguments of Section \ref{sect1} one can expect that some sort of a transition region should exist between a quasi-isothermal torus ''core'' and an infrared-driven part located closer to the source of UV and X-ray radiation. 
The existence of the over-dense region in our 2D gas distributions 
further supports this idea. One can argue that an interesting high density region observed in Figure \ref{fig5},\ref{fig6} can be a sign of a quasi-static/stationary core. Such a region would be a likely place for large scale meridional motions which wrap a quasi-static region.  
The final answer can only be provided by a fully 2.5D time-dependent radiation-hydrodynamical simulations. 
}

\section{Discussion and Conclusions}\label{concl}
We have studied a model of an AGN torus in which obscuration is provided by a radiation-driven wind rather than by a static distribution of gas.
This can occur if the UV and X-ray radiation generated in the inner parts of an accretion disk is reprocessed into the infrared (IR) in the cold, dusty environment 
at approximately 1pc from a supermassive black hole.
We have shown that due to high
dust opacity the pressure of such IR radiation has a profound effect on the torus dynamics and structure. 

Semi-analytic models of a static rotating torus which is supported by infrared pressure on dust grains have been developed by \cite{Krolik07}}. The results confirmed 
that the torus's thickness can be entirely supported by IR radiation, and also raised new questions.  One of the important ones is how to construct a model in which the plasma avoids being blown away (Notice, that critical luminosity 
$L^{\rm UV}_{\rm c,dust} \lesssim 0.1\, L_{\rm edd, e}$), without fine tuning of the parameters?  
In this work we relax the assumption of the static torus and suggest a model which takes into account plasma
motion.

We adopt several approximations and simplifying assumptions. Only constant dust opacity was taken into account, despite the fact that close to the wind (torus)
surface a significant portion of the dust should sublime due to X-ray heating.
In our model we assumed that the wind possesses only a vertical component of the velocity by forcing it to move along cylindrical surfaces. In reality we expect the wind to become more radial at large $r$.

We neglect self-gravitation of the torus despite geometrical and column density arguments which favor torus masses
of  $10^{4} -10^{5}\, M_{\odot}$. It is likely that the self-gravitating instability may operate inside such a torus forming an interacting system of molecular-dusty self- gravitating clouds \citep{KrolikBegelman88, BeckertDuschl04}. 
If there is enough column density and the torus is already geometrically thick it will inevitably intercept
and convert UV and soft X-ray radiation into IR, providing vertical support and possibly suppressing the self-gravitation instability 
\citep{Thompson05,HonigBeckert07}. Thus the optical thickness $\tau_{\rm T}$ of the torus plays important role as an optically thin self-gravitating torus would likely collapse into a thin disk with subsequent star formation \citep{Toomre64}.

In our simplified model, the torus is described by equations of continuous radiation hydrodynamics. 
Even such an oversimplified approach required a complicated numerical treatment. 
The most important part is that in order to obtain the distribution of radiation energy density, $E$ we numerically solve a 2D diffusion equation adopting a flux-limited diffusion approximation.

In our method we solve a simplified system of equations of radiation hydrodynamics, assuming a stationary outflowing wind driven by gradients
of IR radiation pressure.  Our method is not free from serious limitations: we cannot follow ''marginal'' situations, such as a slowly outflowing wind with deceleration. For example, if somewhere in our 2D computational domain such a situation happens, the calculation must stop.  This happens, for example, if the BH luminosity is
too low, $\lesssim 0.1L_{\rm edd}$, or the density is too high, $\tau_{\rm T} \gtrsim  6$ (see Table 1).

{\mybf In most of our simulations
we find three characteristic regions:}
In region I conversion of external UV and soft X-rays into IR provide ample radiation pressure not only to support the torus vertically but to initiate a rigorous outflow;
{\mybf
In region I, the radiation pressure is strong enough to accelerate plasma to velocities $300-400\, {\rm km\,s^{-1}}$ (for $\tau_{\rm T}\simeq 0.5$, 
$\Gamma \gtrsim 0.5$);
At larger $R$ a narrow region II is located where the density rises and the wind is either failed (i.e. first accelerated and then decelerated) or decelerated. }
Region II acts as a barrier separating the dynamical part from the quasi-static one. The latter we call region III and velocities and densities there are small. Region III is quasi-isothermal, although the vertical gradient of the radiation pressure is large enough to 
support its geometrical thickness.

In a real torus, the global flow pattern should be very complex: 
soft X-rays heat the torus surface where cooling cannot compensate for the radiation heating, the temperature rises sharply and outer torus layers start evaporating. Numerical simulations
\citep{Dorodnitsyn08b}, show the formation of a wind with temperature, $T_{\rm w}\simeq10^{4}-10^{7}$K which evacuates $10^{-3}-0.1\,M_{\rm \odot}\,{\rm yr}^{-1}$ from the torus.
In the bulk of the warm absorber flow radiation pressure plays almost no dynamical role. 
Multiple phases of the cold and hot gases may co-exist
in the outflow \citep{KrolikMcKeeTarter81}. In addition to the evaporated gas,
UV-line-driven winds \citep{ProgaStoneKallman00} which are stripped from accretion disk at much smaller radii can also contribute to filling the funnel of the torus. The incident UV and soft X-ray flux is attenuated in this gas.
In the bulk of the torus the gas pressure is much smaller than the
pressure of the infrared radiation which is the major force which keeps the torus geometrically thick.
The
gradients of gas pressure become important in the narrow evaporative layer where temperature jumps from the
cold inner values to the values corresponding to the temperatures of the warm absorber gas, $T_{\rm w}$.

The very high opacity of the cold dusty plasma completely stops
UV radiation somewhere further into the torus, within the narrow layer of the UV photosphere. 
In our current work we identified such a UV photosphere with the inner torus boundary. 
The radiation input was prescribed assuming the distribution of the effective temperature at this boundary. 

Further from the photosphere rotation plays an important role in shaping the density and IR optical depth contours. 
The locally super-critical IR flux creates a radiation-driven outflow. The conditions in such a flow resemble  those in winds of supergiants \citep{LamersCassinelli99} or evolved massive stars \citep{BK-Dor99}.
Hard X-rays,  with energies $E \gtrsim 10$ keV, penetrate much deeper into the torus body 
creating significant local deposition of energy through Compton scattering \citep{Chang_etal07, ShiKrolik08}. 
Some contribution to the radiation field can also come from star formation taking place within the torus or the obscuring flow (Wada \& Norman 2002). 
We will study the influence of these important effects in a future paper.

Close to the torus boundary the infrared flux, ${\bf F}_{\rm IR}\sim -\nabla T$, propagates approximately along the inside normal to the surface. The curvature of the photosphere will significantly influence the distribution of temperature in a thin (of the order of a few mean free paths) thermalization layer. Deeper into the torus, the rotation, the gravitation force, and plasma motion (through the continuity equation) determine
the distribution of $\rho$. At higher heights, $z$ density tends to be lower, and $\nabla T$ is more parallel to $\nabla \tau_{\rm IR}$, where $\tau_{\rm IR}$ is an optical depth. As a result the infrared radiation diffuses in a direction $\sim -\nabla \tau_{\rm IR}$. Further into the torus the radiation tends to make it isothermal. 
Inside these regions there still exists a significant component of the radiation pressure in the $z$ direction, 
but the quasi-static approximation is applicable and the torus is described by models such as those of \citet{Krolik07,ShiKrolik08}. In the intermediate region convective transport of energy may be of importance.

The torus loses mass with an average rate of ${\dot M} \simeq 1 - 10\,M_{\rm \odot}\,{\rm yr}^{-1}$ with negligible kinetic luminosities, $L_{\rm kin}$ which are
$\ll 1\, \% \,L_{\rm bol}$ depending on various model parameters.
This leaves two possibilities: If the gas escapes from the system then the torus will be depleted 
within $10^{4}-10^{5}\,{\rm yr}$ which brings an important connection of the IR-driven obscuration with the AGN feedback problem. Taking into account that radiation-driven flows are believed to be important for the AGN feedback, see e.g. \citep{Begelman04,Fabian10} our results may further favor these ideas.

Yet another possibility is that a considerable part of matter does not leave the torus's potential well, instead forming global vortex-type motions. The impossibility of balancing in one static picture radiation, gravitation and a centrifugal forces is a well known cause of 
meridional flows in rotating radiative stars \citep{Tassoul78,KippenhahnWeigert94}. In a thin accretion disk such imbalance leads to the mass outflow from the disk at a luminosity considerably smaller than the critical one \citep{BKBlinn77}.

{\mybf A high density component in the meridional cut of the velocity distribution is often present in purely hydrodynamical simulations of accretion flows and winds. }
For example, such purely hydrodynamical 2.5D simulations of \cite{Dorodnitsyn08b}, (Figure 5)
reveal the presence of meridional-like, returned current. 
{\mybf Notice that such a region is also present in Figure \ref{fig6} of our simulations as well}.
In these works energy transport is performed by advection. As was shown here, it is entirely plausible that 
$\Pi \gg P_{g}$ in the bulk of the obscuring flow. Thus, inclusion of the infrared radiation pressure and radiative diffusion and advection of the radiative energy density into a time-dependent hydrodynamical framework should demonstrate whether the bulk of the torus is quasi-static and IR supported, 
or in a form of a dusty IR-driven flow.

This research was supported by an appointment at the NASA
Goddard Space Flight Center, administered by CRESST/UMD
through a contract with NASA, and by grants from the NASA
Astrophysics Theory Program 10-ATP10-0171.
We would also like to thank the referee
for constructive comments, which have led to improvement
of the manuscript.

\bibliography{BibList1}

\section*{Appendix: solution of the diffusion equation}
The diffusion coefficient is defined on the augmented grid which is shifted from the $i,j$ grid by $h_{x}$ along the $i$ axes to the left and by
$h_{y}$ down along the $j$ axes.
Thus, with regard to $i,j$ cell we have:
$D^{1}_{i,j}$ is located at the left $i$ boundary; $D^{1}_{i+1,j}$ at the right $i$ boundary; $D^{2}_{i,j}$ at the downside $j$ boundary; $D^{2}_{i,j+1}$ at the upper $j$ boundary. 

When making a time step $\tau$, in the ADI scheme we first perform the inner sweep along $i$ and the outer along $j$ and then 
alternate $i$ and $j$ inner and outer sweeps. First we calculate $E^{*} = E(t+\tau^{*})$, and then $\hat E = E(t+\tau)$.
If the inner sweep is along the $i^{\rm th}$ index, then the finite difference equation to be solved at $i,j$ reads

\begin{equation}\label{fij1}
f_{i,j}=-D^{1}_{i,j}E^{*}_{i-1,j}+\left( D^{1}_{i+1,j} + D^{1}_{i,j} + \frac{h^{2}_{x}}{\tau^{*}}\right)E^{*}_{i,j}-
D^{1}_{i+1,j} E^{*}_{i+1,j} - S^{2}_{i,j}\mbox{,}
\end{equation}
where

\begin{equation}\label{Sij2}
S^{2}_{i,j} = \frac{h^{2}_{x}}{h^{2}_{y}}\left(D^{2}_{i,j+1} ( E_{i,j+1} - E_{i,j}) -  D^{2}_{i,j-1} ( E_{i,j} - E_{i,j-1}) \right)
+E_{i,j}\frac{h^{2}_{x}}{\tau^{*}}\mbox{,}
\end{equation}
where $\tau^{*}=\tau/2$. Then, the inner sweep is made the $j^{\rm th}$ index, and the outer along $j$. The finite difference equation to be solved at $i,j$ reads

\begin{equation}\label{fij2}
f_{i,j} = - D^{2}_{i,j}\hat E_{i,j}+\left( D^{2}_{i,j} + D^{2}_{i,j+1} + \frac{h^{2}_{x}}{\tau^{*}}\right)\hat E_{i,j} -
D^{2}_{i,j+1} \hat E_{i,j+1} - S^{1}_{i,j}\mbox{,}
\end{equation}
where 

\begin{equation}\label{Sij1}
S^{1}_{i,j} = \frac{h^{2}_{y}}{h^{2}_{x}}\left(D^{1}_{i+1,j} ( E^{*}_{i+1,j} - E^{*}_{i,j}) -  D^{1}_{i-1,j} ( E^{*}_{i,j} - E^{*}_{i-1,j}) \right)
+E^{*}_{i,j}\frac{h^{2}_{y}}{\tau^{*}}\mbox{,}
\end{equation}

A sweep method is adopted to solve the resultant tri-diagonal matrix equation via a tridiagonal matrix algorithm \citep{Fedorenko_book}.

A finite difference representation of the boundary conditions (BC) should  preserve
$2^{d}$ order accuracy of the numerical scheme. For example, if the zero flux BC are given at the inner $i$ boundary, we write

\begin{equation}\label{bc1}
(E_{is+1,j}-E_{ is-1,j} )/(2h_{x})=0\mbox{,}
\end{equation}
where $is$ is the first index along $i$, and $is-1$ is the index of the ghost zone. The idea is to express $E_{is-1,j}$ at the ghost zone from the relation for the BC such as (\ref{bc1}), and then to substitute the result into equation (\ref{fij1}) written for the $is$ zone. The resultant equation couples only $is$ and $is+1$ indices:

\begin{equation}\label{fis}
f_{is,j}=\left( D^{1}_{is+1,j} + D^{1}_{is,j} + \frac{h^{2}_{x}}{\tau^{*}}\right)E_{is,j}-
\left( D^{1}_{is+1,j} + D^{1}_{is,j} \right)E_{is+1,j}  - S^{2}_{is,j}\mbox{,}
\end{equation}
where 

\begin{equation}\label{Sis}
S^{2}_{is,j} = \frac{h^{2}_{x}}{h^{2}_{y}}\left(D^{2}_{is,j+1} ( E_{is,j+1} - E_{is,j}) -  D^{2}_{is,j} ( E_{is,j} - E_{is,j-1}) \right)
+E_{is,j}\frac{h^{2}_{x}}{\tau^{*}}\mbox{.}
\end{equation}
Finite difference representations of the boundary conditions at other boundaries are derived in a similar fashion.

\end{document}